\documentclass[12pt]{article} 
\catcode`@=11
\@addtoreset{equation}{section}
\begin{document}
\vskip 50pt
\begin{center}
{\Large \bf  Moment method, Higher order dispersion map and other effects in optical pulse propagation.}
\end{center}
\vskip 20pt
\begin{center}
{\bf {  Basanti Mandal and A.Roy Chowdhury\footnote{E-mail: arcphy@cal2.vsnl.net.in}}\\ 
                           High Energy Physics Division \\      
                             Department of Physics     \\           
                             Jadavpur University        \\     
                              Kolkata - 700032\\
                                    India \\}
\end{center}
\vskip 20pt
[   Analytical and numerical procedures are applied to show that both third and second order dispersion maps can be explicitly constructed and their mutual effects on the optical pulse propagation are analysed. In these connection it is also shown how the other important features such as amplification, intra-channel Raman Scattering(IRS), fibre loss, centre frequency of the pulse spectrum effect the propagation of pulse. Due to the presence of IRS, moment method is adopted which is easily reduced to usual Hamiltonian technique in absence of IRS. It is observed that the long time periodic behaviour is stabilized by the third order dispersion map. Though the introduction of IRS and other effects perturb the situation, one can observe that except for a very special case it  remains stable.]
\newpage
\section{Introduction}
 The study of solitary pulse propagation in optical fibre is nowadays a subject of interest not only to theoritical[1] and experimental physicists[2] but also to technical developers. Due to it's immense posibility for the world of communication, light by light controlled devices and ultra fast spectroscopy, various innovative ideas have been developed and tested. It is now a well known fact that dispersion management[3] plays a central role in the fabrication of long distance fibres for the optimal transmission of the signal. Uptill now the main emphasis is on the second order dispersion effect. A fundamental motivation for the dispersion mapping technique arises from the need to significantly suppress the phase-matched four wave mixing process, which leads to substantial performance penalties in wavelength division multiplexing(WDM) systems[4]. Though dispersion management imposes a strong perturbation to the soliton propagation, original computational evidence of Smith et.al.[5], Gabitov et.al.[6-8] and Golovchenko etal.[9] suggest that solitary wave are indeed robust and can actually cover the long distance. It should be noted that during a given dispersion map period, the pulse can undergo changes in amplitude, width, and chirp. But these changes sampled at appropiate locations of the map can be periodic and thus, a dispersion managed breather is supported.
\par Here we have extended the standard analysis of second order dispersion map to include both second and third order dispersion maps and to monitor the effect of amplitude, IRS, fibre loss and cetre frequency on the characteristics of the pulse propagation. Our method is partly analytic and partly numerical. It is observed from the phase plane dynamic plot that both the second and third order effect can mutually and effectively exist. On the other hand, we have observed that the required long time periodic behaviour of the pulse is altered to some extend when IRS, fibre loss, center frequency are taken care of. A very different phenomenon is observed when the second order dispersion is in the anomalous region and a dispersion map is considered for the third order case, which shows a drastic effect due to IRS and other parameters.
\par Lastly it may be mentioned that since we want to include IRS and other effects which may not be amenable to a variational approach, we have formulated the whole computation by the  moment method.

\section{Formulation: }
We consider the basic equation describing the pulse propagation to be the generalized Nonlinear Schr\"{o}dinger Equation written as $$\frac{\partial {A}}{\partial{z}} +\frac{\alpha }{2}A +\frac{i\beta _2(z) }{ 2} \frac{\partial ^2{A}}{\partial {t ^2}}-\frac{ \beta _3(z)}{ 6} \frac{\partial ^3{A}}{\partial {t^3}} =i\gamma \left(\mid{A}\mid^2A+\frac{i }{\omega_0 }\frac{\partial }{\partial{t}}(\mid{A}\mid^2A)-T_RA\frac{\partial }{\partial{t} }\mid{A}\mid^2\right), \eqno{(1)}$$
where A(z,t) is the slowly varing amplitude of the pulse envelop, $\alpha $ accounts for fibre loss, $\beta _2$ is the group velocity dispersion (GVD), $\beta _3$ is the third order dispersion function(TOD), $\gamma $ is the nonlinear parameter responsibe for self modulation , $w_0$ is the center frequency of the pulse spectrum and $T_R$ represents the Raman parameter responsible for the intrapulse raman scattering(IRS). The basic idea is to treat the optical pulse as a particle with energy E, position T and frequency shift $\Omega$, which are defined via,
$$E=\int_{-\infty } ^{\infty }\mid{A}\mid^2 dt \hskip85 pt  \eqno{(2)}$$
$$T=\frac{1}{E }\int_{-\infty } ^{\infty }t\mid{A}\mid^2 dt \hskip60 pt  \eqno{(3)}$$
$$\Omega=\frac{i}{2E }\int_{-\infty } ^{\infty }\left(A\frac{\partial{A^*} }{\partial{t} }-A^*\frac{\partial{A} }{\partial{t} }\right)dt\eqno{(4)}$$
The root mean square (RMS) width
$$\sigma ^2=\frac{1}{E }\int_{-\infty } ^{\infty }(t-T)^2\mid{A}\mid^2 dt\eqno{(5)}$$
The chirp of the pulse is given as,
$$\tilde {C}=\frac{i}{2E }\int_{-\infty } ^{\infty }(t-T)\left(A\frac{\partial{A^*} }{\partial{t} }-A^*\frac{\partial{A} }{\partial{t} }\right)dt\eqno{(6)}$$

We consider the following trial wave function for the solution of Eq.(1)
$$ A(z,t)= \sqrt{\frac{E\eta}{2}}\;\;\mbox{sech}[\eta(t-T)]  \;\; \mbox {exp}\left[\;i\beta(t-T)^2+i\Omega (t-T)+ i\phi\right],\eqno{(7)}$$
where E(z), $\eta(z)$, $\beta(z)$, $\Omega(z)$, T(z) and $\phi(z)$ are the soliton's  energy, inverse width,  chirp,  frequency, centre  and phase respectively. So, from Eqs.(2)-(6) and using Eq.(1) we get
$$\frac{dE}{dz}= -\alpha E \hskip330 pt  \eqno{(8)}$$
$$\frac{dT}{dz}= -\beta_2\Omega +\frac{\beta_3}{2E}\int_{-\infty } ^{\infty }\mid\frac{\partial A}{\partial t}\mid^2 dt +\frac{3\gamma }{2\omega _0E}\int_{-\infty } ^{\infty }|A|^4 dt \hskip125 pt\eqno{(9)}$$
$$\frac{d\Omega }{dz}= \frac{\gamma T_R}{E}\int_{-\infty } ^{\infty }\left(\frac{\partial |A|^2}{\partial t}\right) dt-\frac{i\gamma }{\omega _0E}\int_{-\infty } ^{\infty }\left(A\frac{\partial{A^*} }{\partial{t} }-A^*\frac{\partial{A}}{\partial t}\right)\frac{\partial |A|^2}{\partial t}dt\hskip75 pt\eqno{(10)}$$   
$$\frac{d\sigma }{dz}= -\frac{\beta_2\tilde {C}}{\sigma }+\frac{\beta_3}{2\sigma E}\int_{-\infty } ^{\infty }(t-T)\mid\frac{\partial A}{\partial t}\mid^2dt\hskip175 pt   \eqno{(11)}$$
$$\frac{d\tilde {C}}{dz}= -\Omega \frac{dT}{dz}-\frac{\beta_2}{E}\int_{-\infty } ^{\infty }\mid\frac{\partial A}{\partial t}\mid^2dt-\frac{i\beta_3}{4E}\int_{-\infty } ^{\infty }\left(\frac{\partial^2A}{\partial t^2}\frac{\partial A^*}{\partial t}-\frac{\partial^2A^*}{\partial t^2}\frac{\partial A}{\partial t} \right)dt\hskip50 pt$$
$$ \left.-\frac{\gamma}{2E}\int_{-\infty}^{\infty}|A|^4dt + \frac{i\gamma }{E\omega _0}\int_{-\infty } ^{\infty }(t-T)\frac{\partial {|A|^2}}{\partial t}  \left(A^*\frac{\partial{A}}{\partial{t}} - A\frac{\partial{A^*}}{\partial t}\right)dt\right.$$
$$\left.+ \frac{i\gamma }{E\omega _0}\int_{-\infty } ^{\infty }\left(A\frac{\partial{A^*}}{\partial{t}} - A^*\frac{\partial{A}}{\partial t}\right)|A|^2dt +
\frac{\gamma T_R}{E}\int_{-\infty } ^{\infty }(t-T)\left(\frac{\partial |A|^2}{\partial t}\right)^2 dt\right. \eqno{(12)}$$
Using the explicit form of A(z,t) given in Eq.(7) and the relations,  $\sigma ^2=\frac{\pi^2}{12\eta^2}$ and $\tilde {C}=\frac{\pi^2\beta}{6\eta^2}$, Eqs. (8)-(12) can be simplified to the following form;
$$\frac{dE}{dz}= -\alpha E \hskip318 pt  \eqno{(13)}$$ 
$$\frac{dT}{dz}= -\beta_2\Omega +\frac{\beta_3}{2}\left[\Omega ^2+(1+\pi^2\frac{\beta^2}{\eta^2})\frac{\eta^2}{3}\right]+\frac{
\gamma  E \eta}{2\omega _0}\hskip140 pt  \eqno{(14)}$$
$$\frac{d\Omega }{dz}=\frac{4}{15}T_R\gamma E\eta^3+\frac{2}{3\omega _0}\gamma \eta E\beta \hskip230 pt  \eqno{(15)}$$ 
$$\frac{d\eta }{dz}=2\beta \eta (\beta _2-\beta _3\Omega )\hskip276 pt  \eqno{(16)}$$
$$\frac{d\beta }{dz}= -\frac{\gamma E\eta ^3}{\pi^2}+2(\beta _2-\beta _3\Omega )(\beta ^2-\frac{\eta ^4}{\pi^2})+\frac{3\gamma E\eta ^3\Omega }{\pi^2\omega _0}\hskip127 pt  \eqno{(17)}$$ 
It is now essential to specify the form of the dispersion management functions $\beta_2(z)$ and $\beta_3(z)$, which will take care of the normal and anomalous dispersion segments.
\par  The dispersion management function  is written as; $$\beta _3(z)= \beta_2(z)=\delta _a +\frac{1}{z_a}\bigtriangleup (\zeta )\eqno{(18)}$$ with dispersion map,
$$\bigtriangleup (\zeta )=\left \{\begin{array} {c}\bigtriangleup _1\; \;  ;\;\;\;\; 0\leq \mid \zeta  \mid < \frac{\theta  }{2}\\ \bigtriangleup_2 \; \;  ;\;\;\;\;  \frac{\theta }{2}\leq \mid \zeta  \mid <\frac{1}{2} \end {array}\right\} $$
where 
$$ \zeta =\frac{z}{z_a}\; ,\;\;\bigtriangleup _1=\frac{2s}{\theta }\; , \;\;\bigtriangleup _2= -\frac{2s}{1-\theta }\;\;\; $$
and map strength, $$s=\frac{\theta \bigtriangleup _1- (1-\theta )\bigtriangleup_2}{4}$$
 One can mention that it is possible to choose different form for $\beta_2$ and $\beta_3$ also. The set of equations(13)-(17) can not be treated numerically and so we take recourse to Runge-Kutta-Felberge method for their solutions.

\section {Numerical Simulation:}
Before going for the complete solution we consider the case of an ideal fibre where $T_R=0$, $\alpha=0.0$ and $\frac {1}{\omega _0}=0.0$. In both the functions $\beta_2$ and $\beta_3$ given in equation (18), we assume $\delta _a=0.0021$ and $z_a=1.0$ where $\delta _a$ is the path average dispersion and $z_a$ is amplifier spacing. The form of $\Delta (\zeta )$ is given in Fig.-1(a). After integration we plot the phase plane diagram in $(\eta,\beta)$ plane and depict the alternative flow through the anomalous and normal regions by different type of lines in Figs.-1(b)-1(e) taking the standard form $\beta_3=0$,$\beta_2\equiv \beta_2(z)$ given in Eq.(18); $\beta_2(z)=\beta_3(z)=$ dispersion function; $\beta_2= -1.0,\; \beta_3(z)=$dispersion function and $\beta_2=1.0,\; \beta_3(z)=$ dispersion function respectively. In last two cases the shape of the phase plane are totally changed. In each of the figures a portion of the figures shows the reversal of the dynamics due to the switching of the sign of the dispersion (in fig-1(b) and 1(c)). Though it is only geometric description, it shows the complete perodic character in absence of $1/\omega _0, \alpha$ and $T_R$. Lastly we continue with the case, $\beta_2=0.0, \beta_3=$dipersion function, depicted in Fig-2. In Fig 2(a) we have the situation with all external effects put to zero and those in fig-2(b) show the closed path in presence of $T_R,\omega _0$ and $\alpha$. which clearly indicates to a stronger effect of TOD(third order dispersion).
\par We next analyse the explicit numerical results for the intensity($\eta$) and chirp($\beta$). In Fig.-3 we exibit the variations of $\eta$ and $\beta$  for the case, $\beta_2=$dispersion function and $\beta_3=0.0$ taking different values of $T_R, \alpha$ and $\omega _0$(values are given in figure caption). The first column of Fig.-3 shows the behaviours of $\eta$ and $\beta$ upto 10 periods of dispersion map and second one exhibits for the single period after traversing nine periods of dispersion map. The long time periodic natures of $\eta$ and $\beta$ are quite evident for the starting situation ($1/\omega _0= \alpha=T_R=0.0$) but change slightly due to the introduction of these effects. We now take up the case when $\beta_2$ and $\beta_3$ both are dispersion maps. For such a case, the variations of $\eta$ and $\beta$ are shown in Fig.-4 and the same conclusion is still valid. One may note that introduction of $\omega _0, \alpha, T_R$ have more influence than in the previous case. Next we consider the case $\beta_2=-1$ and $\beta_3=$ dispersion function. The corresponding situations are shown in Fig.-5.  Here also the starting situation leads to pure long time periodic nature, but remarkable changes are observed with the introduction of $\omega _0, \alpha,T_R$ and due to this fact we have not shown the behaviour after nine map periods.
\par At the end we depict the variations of $\beta$ and $\eta$ in the situation, $\beta_2=0$ but $\beta_3=$ dispersion map. These are exibited in figures 6(a)-6(h). These figures show that even after a longer travelling distance (here we have considered 30 map periods), the pulse shows remarkable stability and the long time periodic behaviour is very much evident, though there may be some deviations from the required features when $\omega_0, \alpha, T_R $ are introduced. This observation gives a further justification for the inclusion of third ordr dispersion map in the construction of fibre.

\section{Conclusion:}
\par Our above analysis suggests that though the basic theory of wave propagation in nonlinear fibre suggests the presence of higher order dispersion and nonlinearity ,for practical situation the effect of such term should be correctly monitored before put to use.  In absence of other extraneous effects the closed trajectories can be observed in phase plane and a map constructed for both second and third order dispersion. The corresponding scenario of chirp and width shows that the system is highly stabilized, yet when IRS and other effects are taken into account and the behaviours of chirp and width show slight fluctuation.

\section{Acknowledgement:} One of the authors (B.M) is grateful to U.G.C. (Govt. of India) for junior research fellowship which made this work possible.
 
\section{References:}

1. {A.Hasegawa and Y. Kodama} - {\it Solitons in optical communications} ,
\par Oxford University Press (Oxford 1995).\\
2. {G.P.Agarwal} - {\it Nonlinear Fibre Optics.} -Academic Press, 
\par San Diego, 2001\\ 
3. {M.J.Ablowitz, G.Biondini, L.A.Ostrovsky} - {\it Chaos.}  {\bf\underline{10}} (2000) 471.\\
4. {D.Anderson}- {\it Phys.Rev.A}  {\bf\underline{27}}(1983) 3135.\\  
5. {N.J.Smith, N.J.Doron, F.M.Knox, W.Forysiak} - {\it Opt.Lett.}  {\bf\underline{21}} (1996) 1981. \\
6. {I.Gabitov, S.K.Turitsyan} - {\it Pisma.V.JETP}  {\bf\underline{63}} (1996) 814. \\
7. {I.Gabitov, E.G.Shapiro, S.K.Turitsyan} - {\it Opt. Comm.} {\bf\underline{134}} (1997) 317.\\ 
8. {I.Gabitov, E.G.Shapiro, S.K.Turitsyan} - {\it Phys. Rev.E.} {\bf\underline{55}} (1997) 3624.\\ 
9. {E.A.Golovchenko, J.M.Jacob, A.N.Pilipetskii, C.R.Menyuk, G.M.Garfer} - 
\par{\it Opt.Lett.} {\bf\underline{22}} (1997) 289.\\

\newpage 
\section{Figure Caption:}
\par Figure-1.(a)Dispersion map.\\
 Figure-1. $(\eta,\beta)$ phase plane diagram with $\frac{1}{\omega _0}=\alpha=T_R=0.0$ and
\par b)  $\beta_2$= dispersion function, $\beta_3$=0.0
 \par    c)  $\beta_2(z) = \beta_3(z)=$ dispersion function.
 \par    d)  $\beta_2 = -1.0, \beta_3=$ dispersion function.
 \par    e)  $\beta_2 = 1.0, \beta_3=$ dispersion function.\\
Figure-2. $(\eta,\beta)$ phase plane diagram for the case, $\beta_2=0.0, \beta_3=$ dispersion function with
\par a) $\frac{1}{\omega _0}= \alpha = T_R = 0.$
\par b) $\frac{1}{\omega _0}=0.33, \alpha = 0.0776, T_R = 1.9$.\\
Figure-3. The variations of $\eta$ and $\beta$ for $\beta_2=$ dispersion function, $\beta_3=0.0$ with
\par (a,e) $\gamma = 1.0, \frac{1}{\omega _0}=\alpha= T_R = 0.0$
\par (b,f) $\gamma = 1.0, \frac{1}{\omega _0}=0.2,\alpha= T_R = 0.0$
\par (c,g) $\gamma = 1.0, \frac{1}{\omega _0}=0.2,\alpha=0.02, T_R = 0.0$
\par (d,h) $\gamma = 1.0, \frac{1}{\omega _0}=0.2,\alpha=0.02, T_R =0.5$\\
Figure-4. The variations of $\eta$ and $\beta$ for $\beta_2=\beta_3=$ dispersion function with
\par (a,e) $\gamma = 1.0, \frac{1}{\omega _0}=\alpha= T_R = 0.0$
\par (b,f) $\gamma = 1.0, \frac{1}{\omega _0}=0.2,\alpha= T_R = 0.0$
\par (c,g) $\gamma = 1.0, \frac{1}{\omega _0}=0.2,\alpha=0.02, T_R = 0.0$
\par (d,h) $\gamma = 1.0, \frac{1}{\omega _0}=0.2,\alpha=0.02, T_R =0.5$\\
Figure-5. The variations of $\eta$ and $\beta$ for $\beta_2=-1.0, \beta_3=$  dispersion function, with
\par a) $\gamma = 1.0, \frac{1}{\omega _0}=\alpha= T_R = 0.0$
\par b) $\gamma = 1.0, \frac{1}{\omega _0}=0.2,\alpha= T_R = 0.0$
\par c) $\gamma = 1.0, \frac{1}{\omega _0}=0.2,\alpha=0.02, T_R = 0.0$
\par d) $\gamma = 1.0, \frac{1}{\omega _0}=0.2,\alpha=0.02, T_R =0.5$\\
Figure-6. The variations of $\eta$ and $\beta$ for $\beta_2=0.0$ , $\beta_3=$dispersion function with
\par (a,e) $\gamma = 1.0, \frac{1}{\omega _0}=\alpha= T_R = 0.0$
\par (b,f) $\gamma = 1.0, \frac{1}{\omega _0}=0.33,\alpha= T_R = 0.0$
\par (c,g) $\gamma = 1.0, \frac{1}{\omega _0}=0.33,\alpha=0.0776, T_R = 0.0$
\par (d,h) $\gamma = 1.0, \frac{1}{\omega _0}=0.33,\alpha=0.0776, T_R =1.9$\\
(Note: The fist and second columns of Figs-3,4 correspond to the behaviours of $\eta$ and $\beta$ upto ten periods of disperison map and single period after travelling nine periods respectively. In first column of Fig.-6 we consider 30 map periods and second column shows 30th period of the map. )

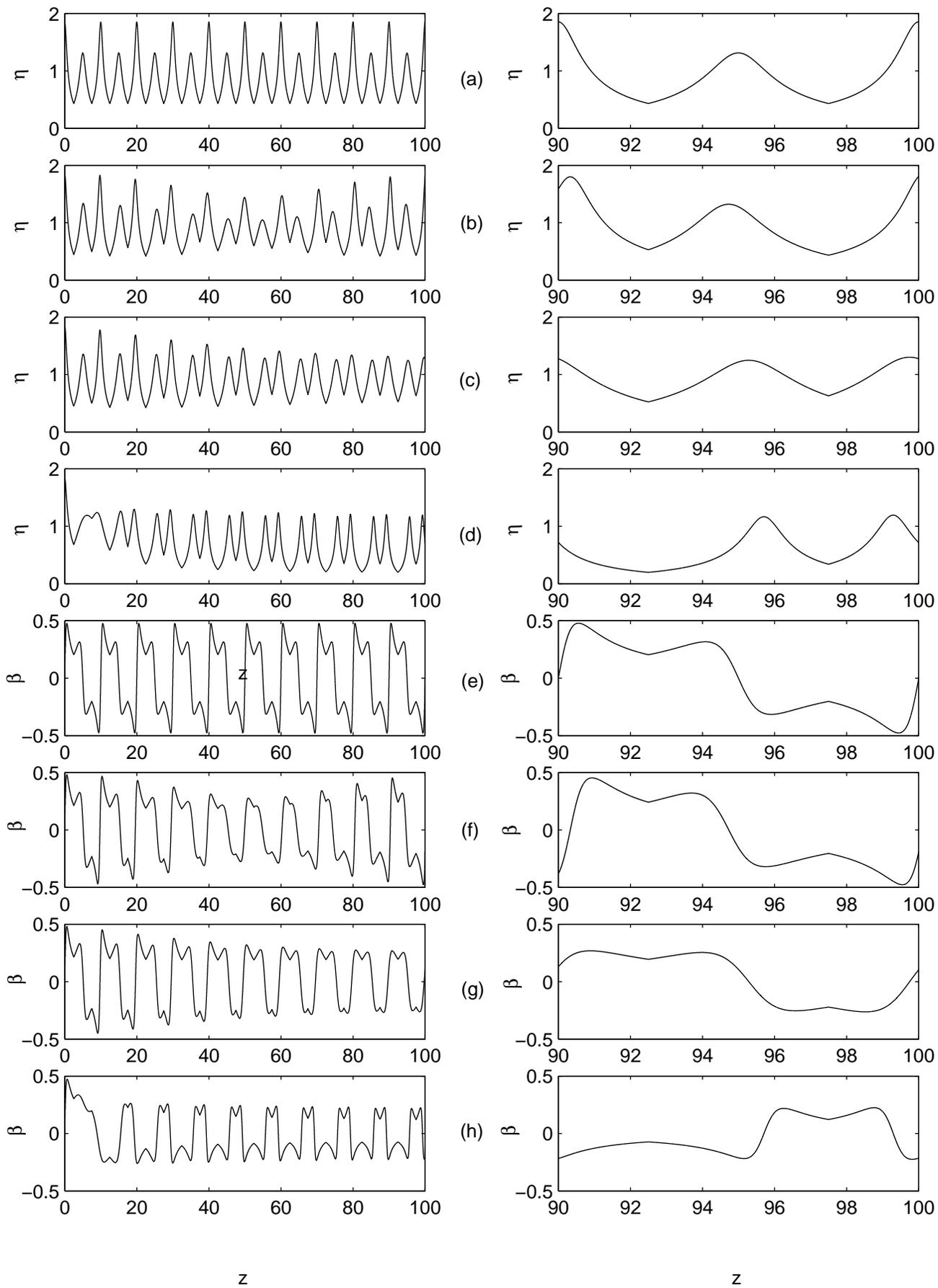
\end{document}